\documentclass[11pt]{amsart}
\usepackage{amsmath,amssymb,mathtools}
\usepackage{graphicx}
\usepackage[margin=1in]{geometry}
\usepackage[hidelinks]{hyperref}
\usepackage{microtype}
\allowdisplaybreaks

\title[Crick--Franklin--Watson Genes of Any Robot]{Defining the Crick--Franklin--Watson Genes of Any Robot\\
(Equals a Finite State Machine) and Defining the Shannon Genetic Code Attached to Each Gene.\\
We Also Define the Krohn--Rhodes Complexity of Any Regular Maximal Prefix Code.}
\author{John Rhodes}
\address{University of California, Berkeley}
\date{August 16, 2026}

\begin{document}
\begin{abstract}
This paper will discuss the relationships among three pillars of mathematics: important research in codes and automata by Marcel-Paul Schützenberger and others; random walks on finite semigroups by Persi Diaconis and others; and advanced techniques from finite semigroup theory used in proving Krohn-Rhodes complexity $c$ is decidable by Stuart Margolis, Anne Schilling, and myself.

Very surprising connections exist between the Fundamental Lemma of Complexity $c$ (epimorphisms between finite semigroups that are one-to-one on subgroups preserve $c$) and coupling from the past in Markov chains, Diaconis' strong stationary time, and the Crick-Franklin-Watson genes of a finite automaton (which will be defined below).
\end{abstract}
\maketitle

\begin{center}
\end{center}

\begin{center}
Conference on Theoretical and Computational Algebra $\cdot$ Guimarães, Portugal $\cdot$ July 1, 2026
\end{center}

\section{Basic Assumptions}

\subsection{Finite-State-Machine Assumption}

Anything---human cells, LLMs, televisions, numerical devices, hurricanes
(von Neumann and Richtmyer \cite{vonneumann-richtmyer1950})---can be modeled arbitrarily closely by a finite state
machine (FSM), or equivalently an automaton.

The starting point is Leibniz's observation that anything can be written
as a sequence of 0's and 1's. Thus everything can be digitized.

This assumption is a fragment of Leibniz's theory of monads and is taken
here as a basic mathematical assumption.

\subsection{Synchronizing Assumption}

Assume that the automaton has a synchronizing sequence \(t\), so that
for every state \(q\) of the automaton,

\[qt = q_{0},\]

where \(q_{0}\) is a fixed state. If the automaton goes wild, put \(t\)
into it---that is, push the ``reset button''---and the automaton returns
to its start state \(q_{0}\).

This is a mild assumption in the real world. For Markov chains in the
mathematical world, it excludes about half of the cases. However, later
in the paper, with ChatGPT's help, we extend the theory to the
non-synchronizing case.

\section{Coupling the Automaton with the Past}

\subsection{Time Convention}

First pass from the automaton to a semigroup, namely the syntactic semigroup of Chomsky--Schützenberger \cite{chomsky-schutzenberger1963}. Choose a direction for time arrows. Here time is taken to run right to
left. If the input alphabet is \(A\), then

\(a_1,\ldots,a_n\in A^+\)

is running backwards in time as

\[\left( \left( ...\left( a_{1}a_{2} \right)a_{3} \right)\cdots a_{n} \right),\]

and the same string runs forwards in time as

\[\left( a_{1}\cdots\left( a_{n - 1}a_{n} \right) \right)...)\]

Of course, these are the same in a semigroup, since

\[(ab)c = a(bc).\]

Thus a semigroup \(S\) with generators \(A\), written \((S,A)\), is a
model of time \cite{rhodes2010}.

\subsection{The Right Karnofsky--Rhodes Expansion}

References for this section are \cite{rhodes-schilling2019,margolis-rhodes-schilling2026}.

Passing from the automaton (the robot) to \((S,A)\), next pass to the
Karnofsky--Rhodes (Kar--R) right expansion

\[(S,A)^{{KR}_{right}}.\]

Here ``right'' means that we are going backwards in time.

Thus, in \((S,A)^{{KR}_{right}}\),

\[a_{1},\ldots,a_{k} \equiv a'_{1},\ldots,a'_{m}\]

if and only if the words \(a_{1},\ldots,a_{k}\) and
\(a'_1,\ldots,a'_m\) are the same in \((S,A)\) and the transition
edges in the Cayley graph of \((S,A)\) along \(a_{1},\ldots,a_{k}\) and
along \(a'_{1},\ldots,a'_{m}\) are the same.

A transition edge \(E\left( q_{1},q_{2} \right)\) is an edge going from
\(q_{1}\) to \(q_{2}\) for which there is no directed path back from
\(q_{2}\) to \(q_{1}\). Intuitively, a transition edge
\(E\left( q_{1},q_{2} \right)\) is a ``big event'' occurring in the
past: at \(q_{1}\) the big event has occurred, while at \(q_{2}\) it has
not yet occurred. The time direction is known \(\rightarrow\) unknown.

A non-transition edge \(E\left( q_{1},q_{2} \right)\) means that not
much has changed from \(q_{1}\) to \(q_{2}\).

\subsection{Genes}

The semigroup \((S,A)^{{KR}_{right}}\) will be called the genes of the
initial automaton, or robot. Thus the genes of the robot go into the
past and remember big events.

Now \((S,A)^{{KR}_{right}}\) is the right Cayley graph of a semigroup
with generators \(A\) whose minimal ideal is

\[K = LZ(n) = \{ 1,\ldots,n\},\quad\quad ab = a,\]

called the genes. Therefore we can act on the left.

So time now goes forward from \(K = LZ(n)\). Thus

\[\left( (S,A)^{{KR}_{right}},\ LZ(n) = genes \right)\]

is going forward in time. A gene \(a_{1}\cdots a_{n}\) is a word for
which \(a_{1}\cdots a_{n}\) is a constant map, equivalently a word in
the minimal ideal \(K\).

The genes go into the past and remember big events; the left action on
the genes is therefore going forward in time. The right
Karnofsky--Rhodes expansion \((S,A)^{{KR}_{right}}\) acts to the left on
the genes.

\section{Environment and Shannon Code of a Gene}

\textbf{Definition 3.1 (First return strings and Shannon code).} For
each gene \(g\), define \(R_{g}\) to be the set of all strings \(t\in A^+\) such that

\[tg = g,\]

but, for every proper suffix \(t'\) of \(t\),

\[t'g \neq g.\]

This set \(R_{g}\) of first return strings relative to \(g\) is a
regular suffix-maximal code; regular means, it has a Kleene expression
and is accepted by a finite state machine. \(R_{g}\) is called the
\emph{Shannon code} of the gene \(g\).

By a standard theorem in Markov chains, using sums of independent
variables, the expected length of the prefix code obtained by reversing
the strings (notice this change) \(R_{g}\) is \(1/\Pr(g)\), where
\(\Pr(g)\) is the probability of the gene given by the stationary
distribution of the Markov chain of A acting on the left of the genes.

\textbf{Definition 3.2 (Krohn--Rhodes complexity of a prefix code).} The
Krohn--Rhodes complexity of a regular maximal prefix code is defined to
be the Krohn--Rhodes complexity \(c\) of the reduced complete
deterministic flower automaton of \(rev\left( R_{g} \right)\),
consisting of all loops at one state \(q_{0}\). See the book \emph{Codes and Automata} \cite{berstel-perrin-reutenauer2010} for details.

Since it is known, the flower automaton \(E_g\), based at \(g\), is the largest refinement of the two-set partition of the genes into \(g\), and the rest of the genes, which is a congruence. This implies our automata acting on genes is the subdirect product of the flower automaton, so it follows that \(c\) of our starting automata is the maximum of \(\{c f_g \mid g \text{ is a gene}\}\).

\subsection{A One-State Automaton}

Let \(g\) be the only state of the automaton. Let \(A\) be the alphabet
and let \(p\) be a positive probability distribution on \(A\). Again,
\(p(a)\) is the probability that \(a\) occurs in the environment.

The right Karnofsky--Rhodes construction, with left action, has genes
\(A\) and forward left action by \(a\) is given by a.a1 = a.

\emph{Typewriter interpretation.} This automata is a typewriter with
genes as the keys: hit key \(h\), and \(h\) comes out. Suppose, for
example, that

\[p(h) = \frac{5}{100}.\]

Even in this trivial situation, Markov-chain theory proves that on
average one must wait

\[\frac{100}{5} = 20\]

times to see the next \(h\).

In general, the average length of a string in \(R_{g}\) with respect to
\(p\) is one over the probability that \(g\) occurs in the unique
stationary distribution on the genes forced by \(p\).

\subsection{A Two-State Example}

Let the states be \(\{ 1,2\}\). Let the right-action inputs be

\[A = \{ a,b\},\]

with equal probabilities \(1/2\). Let

\[1 \cdot a = 2 \cdot a = 1,\]

so that \(\cdot a\) is constant \(1\). Then \(\cdot b\) flips the
states, so

\[1 \cdot b = 2,\quad\quad 2 \cdot b = 1,\]

and

\[S = \langle \cdot a, \cdot b\rangle\]

be the semigroup generated by \(\cdot a\) and \(\cdot b\). Then
\(\left( S,\{ a,b\} \right)\) is the semigroup of all maps on
\(\{ 1,2\}\) acting on the right.

The right Cayley graph of the right Karnofsky--Rhodes expansion

\[\left( S,\{ a,b\} \right)^{{KR}_{right}} = \left( T,\{ a,b\} \right)\]

has order \(5\).

In Figure 1 (in Section 4. below), the label \(t\) denotes a transition edge.

\begin{figure}[ht]
\centering
\includegraphics[width=0.88\textwidth]{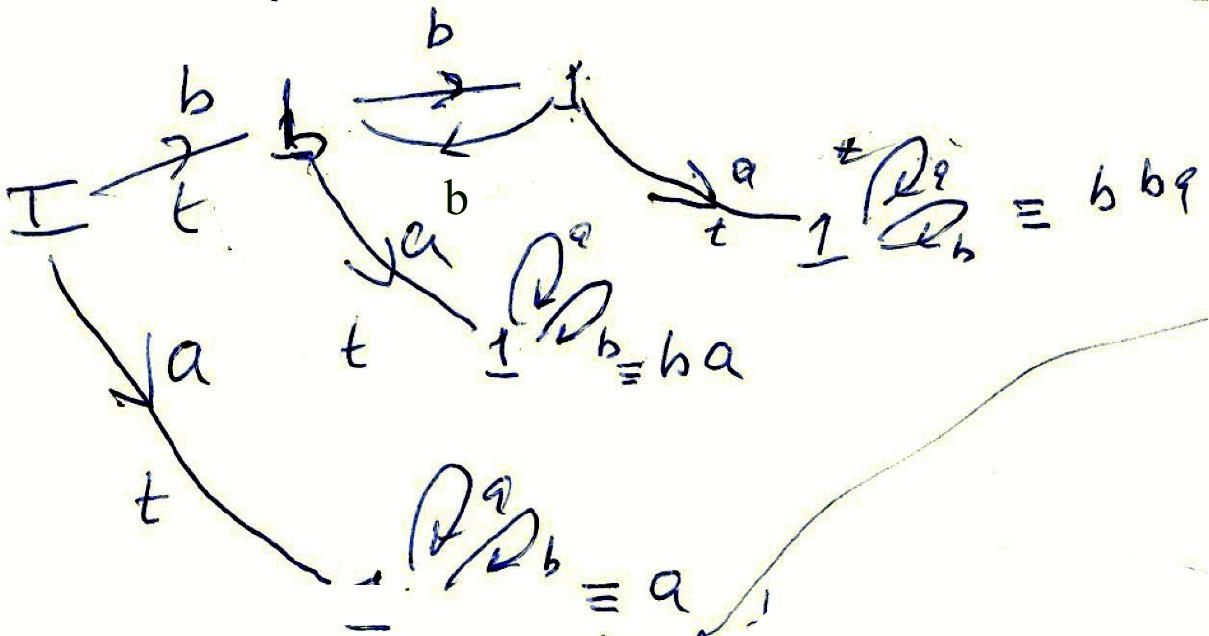}
\caption{Sketch of the right Cayley graph for the two-state example.}
\end{figure}

The three genes are

\[b,\quad\quad ba,\quad\quad bba.\]

The left action on the genes by
\(b.\ and\ a.\quad is\quad a.\ is\ constant\ a,\)

and \(b.\ \) is b.a = ba, b.ba = bba and b.bba = ba. The stationary
probability distribution st( ) on the genes stemming from a and b having
equal probabilities is st(a) = ½, st(ba) = 2/6 and st(bba) = 1/6

\section{Details of Coupling from the Past}

References for this section: \cite{diaconis-fill1990,aldous-diaconis1986}.

We consider time going from right to left,

\[\leftarrow .\]

Given a semigroup \(S\) with generators \(X\), consider \((S,X)\)

right Karnofsky--Rhodes, so that we are going backwards in time,

\[\rightarrow .\]

Given \(x_{1},\ldots,x_{k} \in X\), first consider

\textbf{(1)}

\[\left( I = s_{0} \right),\quad\quad x_{1} = s_{1},\quad\quad x_{1}x_{2} = s_{2},\quad\quad\ldots,\quad\quad x_{1}\cdots x_{k} = s_{k},\]

moving backwards in time in the right Cayley graph of \((S,X)\). Then
add the arrows of the right Cayley graph:

\[I \rightarrow^{x_{1}}I \cdot x_{1} = s_{1},\]

\[s_{1} \rightarrow^{x_{2}}s_{1} \cdot x_{2} = x_{1}x_{2},\]

\[\vdots\]

\textbf{(2)}

\[s_{k - 1} \rightarrow^{x_{h}}s_{k - 1} \cdot x_{k} = s_{k} = x_{1}\cdots x_{k}.\]

Omit the arrows \(s \rightarrow^{x}s \cdot x\) when \(sRs \cdot x\),
that is, when the edge is \textbf{not} transient. The remaining arrows
are simply called arrows.

Then write:

\textbf{(3)}

\[{KR}_{right}\left( x_{1},\ldots,x_{k},\alpha \right) = \left( (\alpha)Arrows,\alpha_{S} \right),\quad\quad\alpha \in S.\]

For \(\alpha_1,\alpha_2\in X^+\), the product is written as

\textbf{(4)}

\[\left( \left( \alpha_{1} \right)Arrows,\left( \alpha_{1} \right)_{S} \right) \cdot \left( \left( \alpha_{2} \right)Arrows,\left( \alpha_{2} \right)_{S} \right) = \left( \left( \alpha_{1} \right)Arrows \cup \left( \alpha_{1} \right)_{S} \cdot \left( \alpha_{2} \right)Arrows,\left( \alpha_{1}\alpha_{2} \right)_{S} \right),\]

where \(\left( \alpha_{1} \right)_{S} \in S\) acts on the left on
\(\left( \alpha_{2} \right)Arrows\).

The action on an arrow is

\textbf{(5)}

\[s\cdot(s_1\xrightarrow{x_j}s_1x_j)=\]

\begin{itemize}
\item
  \(ss_1\xrightarrow{x_j}ss_1x_j\), if \(ss_1\) is not \(\mathcal R\)-related to \(ss_1x_j\); and
\item
  if \(ss_1\) is \(\mathcal R\)-related to \(ss_1x_j\), and omitted otherwise.
\end{itemize}

Thus \((S,X)\) right Karnofsky--Rhodes is a subsemigroup of a right
semidirect product of a semilattice with \(S\).

To complete the coupling from the past, act on the left. Notice the
right--then--left pattern.

This is like Munn's construction for free inverse semigroups, and like
the Margolis--Meakin construction for Rabin tree automata \cite{munn1955,margolis-meakin-graph1993,margolis-meakin-trees1993}.

\section{Diaconis Strong Stationary Time and Rhodes--Schilling Synchronization Theory}

Bingo! Bingo! We now show how the SST (strong stationary times) of
Diaconis and Fill \cite{diaconis-fill1990} and Aldous and Diaconis \cite{aldous-diaconis1986} are implied by
(S,X) right Karnofsky-Rhodes acting left.

\textbf{Proposition.} Let \(S\) be a finite semigroup with generators
\(X\), written \((S,X)\), and assume that the minimal ideal in \(S\) is
zero. Pass to \((S,X)\) right Kar--R acting on the left on its minimal
ideal \(LZ(q)\). The elements \(g\) of this minimal ideal are the genes.
Then this left action satisfies the Strong Stationary Time (SST) of
Diaconis.

\emph{Proof.} The proof follows from Diaconis' SST together with
Rhodes--Schilling unified synchronization theory \cite{rhodes-schilling2019}. After (S,X) right KR is applied, RS
proved the stationary distribution is given by the probability of all
strings first hitting a gene g and the other conditions of SST.

Using the 1/2 \(\ell_{1}\)-norm for mixing time, and using
\(\phi\) for the random variable that records the first time hitting the
minimal ideal---that is, the genes of \(S\)---we have

\[mix(t) \leq \Pr(\phi > t).\]

If \(E = E\lbrack\phi\rbrack\) is finite, then Markov's inequality gives

\[mix(t) \leq \frac{E}{t + 1}.\]

Also, acting on the left on the genes, the stationary distribution
\(st(g)\) is the sum of the probabilities of all strings going
right---hence backwards in time---and first hitting \(0\). This is
essentially the Rhodes--Schilling theory in the synchronization case.
Note that it does not change when the time direction is changed. This
completes our outline of the proof.

There are thirteen ways to compute \(E\). (See \cite{rhodes2027}.) One way is that \(E\) is the
sum of the probabilities of strings that do not hit the minimal ideal.
Another way is Feller's overlap theorem \cite{feller1968} and \cite{berstel-perrin-reutenauer2010}.

See \cite{berstel-perrin-reutenauer2010}. This
follows the philosophy of adding more finite semigroup theory to
Diaconis' theorems. See the forthcoming \cite{rhodes2027}.

\subsection{Chat Extension to the General Finite Markov Chain Case}

\emph{Note to the reader:} This is just the proceeding of the talk given at TCA Guimarães on 1 July 26. This section needs more expansion, which will be done in a future paper. So for now, this Subsection 5.1 should be considered a conjecture.

Given \((S,X)\), where \(S\) is arbitrary finite with generators \(X\),
and a probability distribution \(p\) on \(X\), let the Rees matrix
completely simple semigroup be the min ideal of S denoted

\[M\lbrack G,A,B,C\rbrack,\quad\quad C:B \times A \rightarrow G.\]

We can assume \(S\) is GM, meaning that it acts faithfully on the left
and right of \(I\), by lumping. See \cite[Chapter 4]{rhodes-steinberg2009}.

To keep the reader attentive, now let the time arrow go left to right,

\[\rightarrow ,\]

and consider (T,X) = \((S,X)\) left Karnofsky--Rhodes, and then act on
the right side.

The new minimal ideal, or kernel, is

\(K = M\left\lbrack G,A,B^{'}C^{'} \right\rbrack.\)

The \(L\)-class \(B\) grows to \(B' \rightarrow B\), and if
\({b'}_{1}\ and\ {b'}_{2}\ map\ to\ the\ same\ element\ of\ B\), then
\({b'}_{1}\) and \({b'}_{2}\) are left proportional by \(G\).

The stationary distribution \(d\) on \(A \times G \times B'\) is the
reverse lumping of the stationary distribution on

\[U = A \times G \times B.\]

By Berstel, Perrin, and Reutenauer \cite{berstel-perrin-reutenauer2010} on
\(K = A \times G \times B'\), the stationary distribution is the product
of the distribution on \(LLM(S)\), the distribution on \(RLM(S)\), and
the uniform distribution \(1/|G|\) on \(G\). This is denoted by \(\pi\).

Thus \(LLM(T)\) with kernel \(LZ(T)\) and \(RLM(T)\) with kernel
\(RZ(T)\) come from the synchronized Rhodes--Schilling case.

The proposed upper bound for the \(\frac12\ell_1\)-metric mixing
time is

\(mix(t) \leq \Pr(\phi > t) + 2d(t),\) denoted (*)

where \(d(t)\) is the mixing time of a random variable \(H\) on
\((G,X)\) depending on \((T,X)\).

For \(x \in X\), first define

\[\left( a_{0},g,b' \right) \cdot x = \left( a_{0}.g \cdot (b')x,b' \cdot x \right),\]

and then define, for each generator x, the density of the random
variable on B\('\) given by b\('\) to (b\('\))x

denoted (g)D\textsubscript{x} = sum of all (b\('\))x = g and finally let
(g)H = sum of p(x).(g) D\textsubscript{x} over all x.

Thus H gives a random walk on \((G,X)\), with mixing time \(d(t)\).

\emph{Proof sketch.} We have SST to \(B'\), which gives the first term.

Then the sum on \(a_{0} \times G \times \underline{B}\) does not depend
on \(a_{0}\).

Now assume at time t the g coordinate is off by t1 in {[}0,1) and the b'
coordinate is off by t2in {[}0,1).

Then (½(g +t1)(b' + t2)) - 1/2gb' = ½(gt1 + b't2 + t1t2) \textless= ½(
2t1 + t2) = t1 +1/2t2 so (*) follows.

By reverse lumping, we can get back to \(B\), so \(d(t)\) is the group
correction term. \(\square\)\\
\strut \\
\textbf{Example 4:}
\(\mathbf{G}\mathbf{=}\mathbf{C}_{\mathbf{5}}\)\textbf{, lazy
nearest-neighbor group walk}

Let

\[G = C_{5} = \{ 0,1,2,3,4\}.\]

Let

\[B = \{ - 1,0,1\}.\]

Suppose the \(B\)-chain has stationary distribution

\[d_{B}( - 1) = d_{B}(0) = d_{B}(1) = \frac{1}{3}.\]

Assume the group increment is

\[\theta(b,x) = b,\]

so the group increment is \(- 1\), \(0\), or \(1\), each with
probability \(1/3\). Then

\[f( - 1) = f(0) = f(1) = \frac{1}{3}\]

on \(C_{5}\).

So the induced group walk is the lazy nearest-neighbor walk on the
5-cycle:

\[g \mapsto g - 1,\quad g,\quad g + 1,\]

each with probability \(1/3\).

The Fourier coefficients are

\[\widehat{f}(j) = \frac{1}{3}\left( 1 + e^{2\pi ij/5} + e^{- 2\pi ij/5} \right) = \frac{1}{3}\left( 1 + 2\cos\frac{2\pi j}{5} \right).\]

For \(j = 1,2,3,4\),

\[\widehat{f}(1) = \widehat{f}(4) = \frac{1}{3}\left( 1 + 2\cos\frac{2\pi}{5} \right),\]

\[\widehat{f}(2) = \widehat{f}(3) = \frac{1}{3}\left( 1 + 2\cos\frac{4\pi}{5} \right).\]

Numerically,

\[\left| \widehat{f}(1) \right| = \left| \widehat{f}(4) \right| \approx 0.5393,\]

\[\left| \widehat{f}(2) \right| = \left| \widehat{f}(3) \right| \approx 0.2060.\]

Thus

\[d_{G}(s) \leq \frac{1}{2}\left( 2(0.5393)^{2s} + 2(0.2060)^{2s} \right)^{1/2}.\]

So

\[d_{W}(t) \leq \min_{0 \leq s \leq t}\left\lbrack P_{I}(t - s) + \frac{1}{2}\left( 2(0.5393)^{2s} + 2(0.2060)^{2s} \right)^{1/2} \right\rbrack.\]

For large \(s\), the dominant group term is approximately

\[d_{G}(s) \lesssim \frac{1}{\sqrt{2}}(0.5393)^{s}.\]

So the whole semigroup walk mixes once both things have happened:

\[\text{fallen into }I\]

and

\[\text{the lazy walk on }C_{5}\text{ has mixed.}\]

\medskip
\noindent\textbf{Example 5: $G=C_m$, uniform $\{-1,0,1\}$ walk}\par
\smallskip

More generally, take

\[G = C_{m},\]

and suppose the induced group law is

\[f( - 1) = f(0) = f(1) = \frac{1}{3}.\]

Then

\[\widehat{f}(j) = \frac{1}{3}\left( 1 + 2\cos\frac{2\pi j}{m} \right).\]

Therefore

\[d_{G}(s) \leq \frac{1}{2}\left( \sum_{j = 1}^{m - 1}\left| \frac{1 + 2\cos(2\pi j/m)}{3} \right|^{2s} \right)^{1/2}.\]

The semigroup mixing bound is

\[d_{W}(t) \leq \min_{0 \leq s \leq t}\left\lbrack P_{I}(t - s) + \frac{1}{2}\left( \sum_{j = 1}^{m - 1}\left| \frac{1 + 2\cos(2\pi j/m)}{3} \right|^{2s} \right)^{1/2} \right\rbrack.\]

For large \(m\), the dominant Fourier mode is \(j = 1\), and

\[\widehat{f}(1) = \frac{1}{3}\left( 1 + 2\cos\frac{2\pi}{m} \right) \approx 1 - \frac{4\pi^{2}}{3m^{2}}.\]

So the group mixing scale is of order

\[m^{2}.\]

Thus in this example,

\[\text{mixing time of }W\]

is controlled by

\[\max\left\{ \text{time to fall into }I,\quad m^{2} \right\}.\]

\section{Memory and Semigroup Stabilizers}

The references for this section include \cite{lesaec-pin-weil1991} (for LeSaec-Pin-Wil) and \cite{tilson1971,tilson1974,tilson1976} (for Tilson).

\subsection{Right Stabilizers}

If \(S\) is a finite semigroup, define the right stabilizer of \(s\) by

(s)\({st}_{S} = \{ s_{1} \in S:ss_{1} = s\} \leq S.\)

From Tilson, we have the following proposition.

\textbf{Proposition 7.1.} The right stabilizers of \((S,A)\) left
Karnofsky--Rhodes are aperiodic and form an \(L\)-chain in \(S\), and
therefore are \(R\)-trivial.

From Le Saec--Pin--Weil, we have the following proposition.

\textbf{Proposition 7.2.} If \(Z_{p}\), for sufficiently large \(p\) depending on an automaton or on its semigroup,

and d the J length of the semigroup is used to replace by

\[(Z_p)^d\mathbin{\wr} sg(A),\]

then all right stabilizers become idempotent, that is, a band.

Thus, if \(sg(A)\) is first replaced by left Karnofsky--Rhodes with
respect to its generators, then doing

Proposition 7.2 above yields a right stabilizer that is a band of
idempotents forming an \(L\)-chain

\[\{ e_{1},\ldots,e_{k}\},\quad\quad e_{i} = e_{i}^{2},\]

with the relation indicated in the source for \(i \leq j\).

\textbf{Conjecture (Junk DNA).} The null elements in the left stabilizer
of \((S,A)\) right Kar--R are the ``junk'' DNA on which the results are
written, as the genes can produce more information than they have space
to write, except for the junk DNA.

\textbf{Remark.} \emph{Someone will win a Nobel prize for this someday.}

\section*{Appendix A. Every Finite Markov Chain Is a Random Walk on a Finite Semigroup}

Let \(M\) be an \(n \times n\) right-action stochastic matrix; that is,
\(M(i,j) \in \lbrack 0,1\rbrack\) and the rows sum to \(1\). Think of
one frog \(M\) jumping around on \(n\) petals.

There exists a total self-map \(f\) on \(n\) letters such that

\[M\left( i,(i)f \right) \neq 0,\quad\quad i = 1,\ldots,n.\]

If

\[p_{f} = \min_{i}M\left( i,(i)f \right),\]

then

\[M - p_{f}(f)_{RM}\]

is a matrix with coefficients in \(\lbrack 0,1\rbrack\) whose rows all
sum to some \(c \in \lbrack 0,1\rbrack\). Here \((f)_{RM}\) is the
right-monomial matrix corresponding to \(f\).

By continuing, one can write

\[M = p_{f_{1}}\left( f_{1} \right)_{RM} + \cdots + p_{f_{m}}\left( f_{m} \right)_{RM},\]

where \(p_{f_{1}},\ldots,p_{f_{m}}\) is a positive probability
distribution \cite{birkhoff1946}. Let

\[S = \langle f_{1},\ldots,f_{m}\rangle\]

be the semigroup generated by the \(f_{i}\), with
\(A = \{ f_{1},\ldots,f_{m}\}\). This proves the Birkhoff/von Neumann Theorem for right stochastic square matrices \cite{birkhoff1946,vonneumann1953}.

There are two random walks on \((S,A)\). The first is on the right
regular representation, equivalently the right Cayley graph of
\((S,A)\). The states are \(S^{I}\), and the action is

\[s_{i} \mapsto s_{i}a,\quad\quad I \mapsto a.\]

The stationary distribution, denoted \(dist\), is positive exactly on
\(K\), the kernel or minimal ideal of \(S\). This kernel is a completely
simple Rees matrix semigroup

\[K = M\lbrack G,A,B,C\rbrack,\]

where \(G\) is a finite group, \(A\) and \(B\) are nonempty sets, and

\[C:B \times A \rightarrow G.\]

This Markov chain starts at \(I\) and has

\[\Pr\left( a = f_{i} \right) = p_{i}.\]

This is the right semigroup version.

The classical right version is obtained by choosing a start state
\(q \in K\),

\[q = (a,g,b),\]

but the right Markov chain depends only on \((g,b)\).

We are interested in \(dist\), which is the same in both cases, and in
the upper bounds on the mixing times, which in general are different in
the two cases.

\end{document}